# A Versatile Side Entry Laser System for Scanning Transmission Electron Microscopy

*Ondrej Dyck[1], Olugbenga Olunloyo[2], Kai Xiao[1], Benjamin Wolf[3], Thomas M. Moore[3], Andrew R. Lupini[1], Stephen Jesse[1]*

[1] *Center for Nanophase Materials Sciences, Oak Ridge National Laboratory, Oak Ridge, TN*
[2] *Department of Physics and Astronomy, University of Tennessee, Knoxville, TN*
[3] *Waviks Incorporated, Dallas, TX*

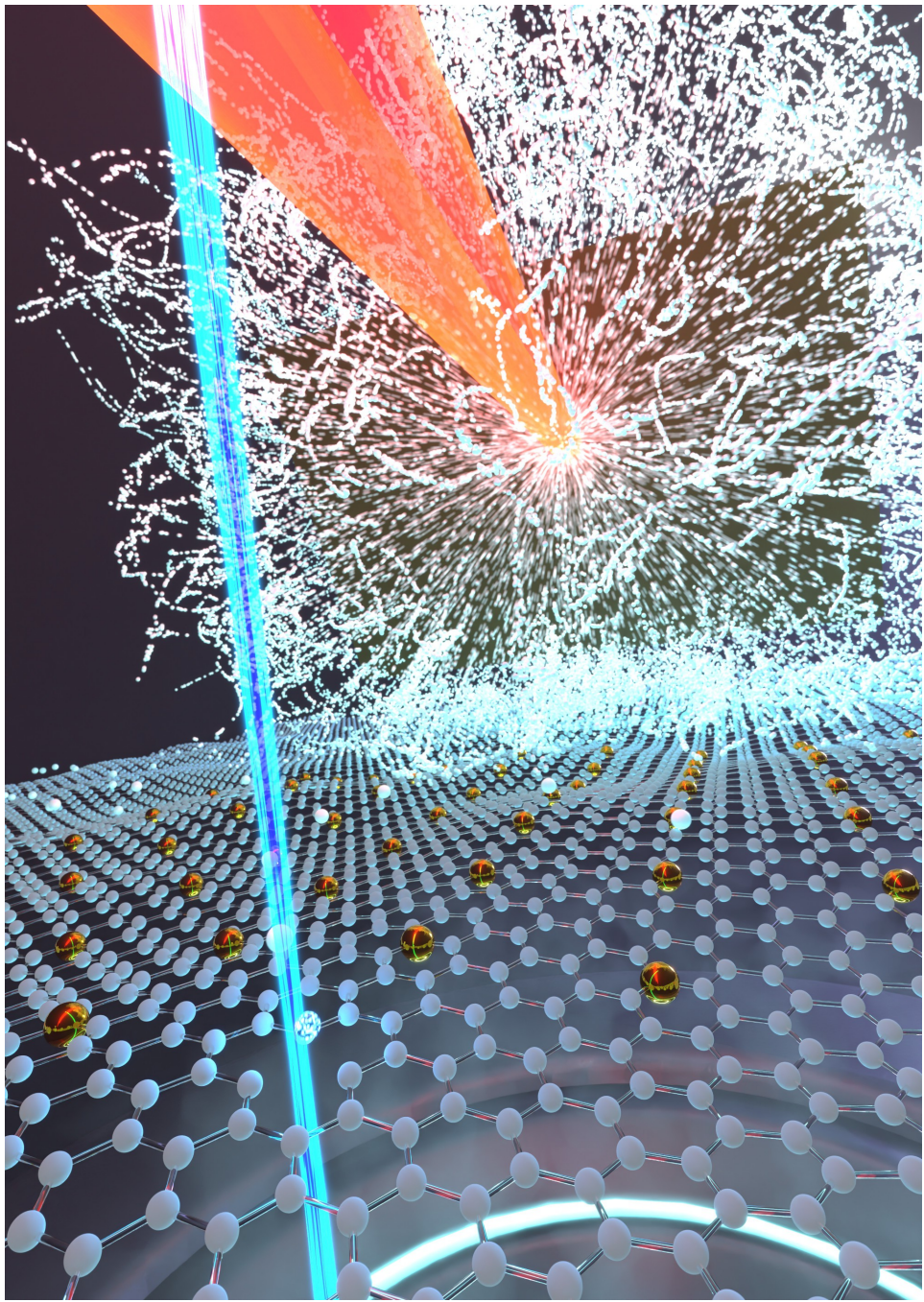

**Abstract**

We present the design and implementation of a side entry laser system designed for an ultra-high vacuum scanning transmission electron microscope. This system uses a versatile probe design enclosed in a vacuum envelope such that parts can be easily aligned, modified, or exchanged without disturbing the vacuum. The system uses a mirror mounted on the sample holder such that the sample can be illuminated without being tilted. Notably the mirror can be removed and replaced with an ablation target and a higher power laser used to ablate material directly onto the sample. We argue that new capabilities hold the potential to transform the electron microscope from an analysis tool towards a more flexible synthesis system, where atomic scale fabrication and atom-by-atom experiments can be performed.

**Introduction**

A significant challenge for the fabrication and examination of emergent quantum properties of materials that arise from local structural defects, is the lack of strategies available for experimental intervention. In an idealized experiment, one variable is changed while all others are held fixed to enable observation of the unambiguous response to that change. The variety of interventions—control over variables—enable a methodical search through, and empirical documentation of, the system response. In the case of extreme localization, where quantum properties naturally emerge, there is often a lack of controllability over the creation, alteration, and detection of the systems of interest. Theorists can easily place every atom where they want and systematically vary parameters at will to explore (theoretically) emergent material properties. Experimentalists are circumscribed by the complexity of the real world and rely heavily on technologically innovative strategies to advance their ability to exert control. To accelerate discovery, a tight loop between experimental design, measurement, interpretation, and subsequent experiment is needed. Currently, however, there are few ways to interact with localized structural elements to design new experiments at the scales relevant for quantum effects to emerge and be explored.

A recent publication by some of the present authors explored the idea of performing synthesis processes within the vacuum chamber of a scanning transmission electron microscope (STEM).[57] Dubbed the "synthescope", this approach would enable direct, atomically resolved imaging of growth processes as well as incorporation with a steadily expanding set of electron beam (e-beam) induced transformations that operate on a local level, for example, the movement,[1] insertion,[2–6] and patterning[1,7] of impurity atoms or holes[8] and the carving of 2D materials[9–14] (for an extended list of examples see reference,[15] especially the supplemental table). Motivation for the development of the synthescope idea centers around the concept that the beam-sample interaction volume furnishes a chamberless synthesis environment (CSE) within which reactions—in general—can occur, shown in Figure 1(a). From this point of view, e-beam induced deposition (EBID) is a specific reaction that can occur within the CSE. This specific reaction involves the beam-induced dissociation (cracking) of a precursor molecule and the subsequent chemical bonding of the molecular fragment to the surface of the substrate, illustrated in Figure 1(b). The fundamental limit, i.e. the smallest amount of material that can be deposited, is the molecular fragment, illustrated in Figure 1(c). This limit was achieved by van Dorp et al. in 2012[16] where they were able to demonstrate molecule-by-molecule deposition on graphene using a $W(CO)_6$ precursor molecule in a STEM.

A different approach was taken by several of the present authors, in Dyck at al. where,[2] instead of dissociating the precursor, the substrate was chemically altered (damaged) by e-beam irradiation, allowing single atoms to be incorporated into the substrate (graphene). This process is illustrated in Figure 1(d) and again the fundamental limit is a single atom, shown in Figure 1(e). This process was shown to be applicable to a variety of elements[4–6] and recognized as a variant of a more generalized EBID concept. Figure 1 attempts to illustrate the parallels between conventional EBID (where precursors are dissociated to drive the reaction) and atomic deposition (where the substrate is modified to drive the reaction), and to highlight how the two different strategies can be conceptually unified.

If we now relinquish the specification of any particular reaction, it becomes apparent that many different types of growth and synthesis are possible. Many different reactions could occur, and we seek to elucidate the parameters that govern the products by controlling the local environmental factors.

In an attempt to develop these capabilities and support the exploration of a generalized CSE approach to nanoscale and atomic manipulation and fabrication technologies, a conceptual change in perspective of the microscope itself must take place. This concept was laid out in a recent publication where the term "synthescope" was used to emphasize the synthesis aspects of this view of the microscope.[17] To provide a realistic pathway toward the accomplishment of such a vision, two plausible embodiments—conceptualized to require minimal modification to the existing microscope hardware—were presented, depicted in Figure 1(f) and (g).

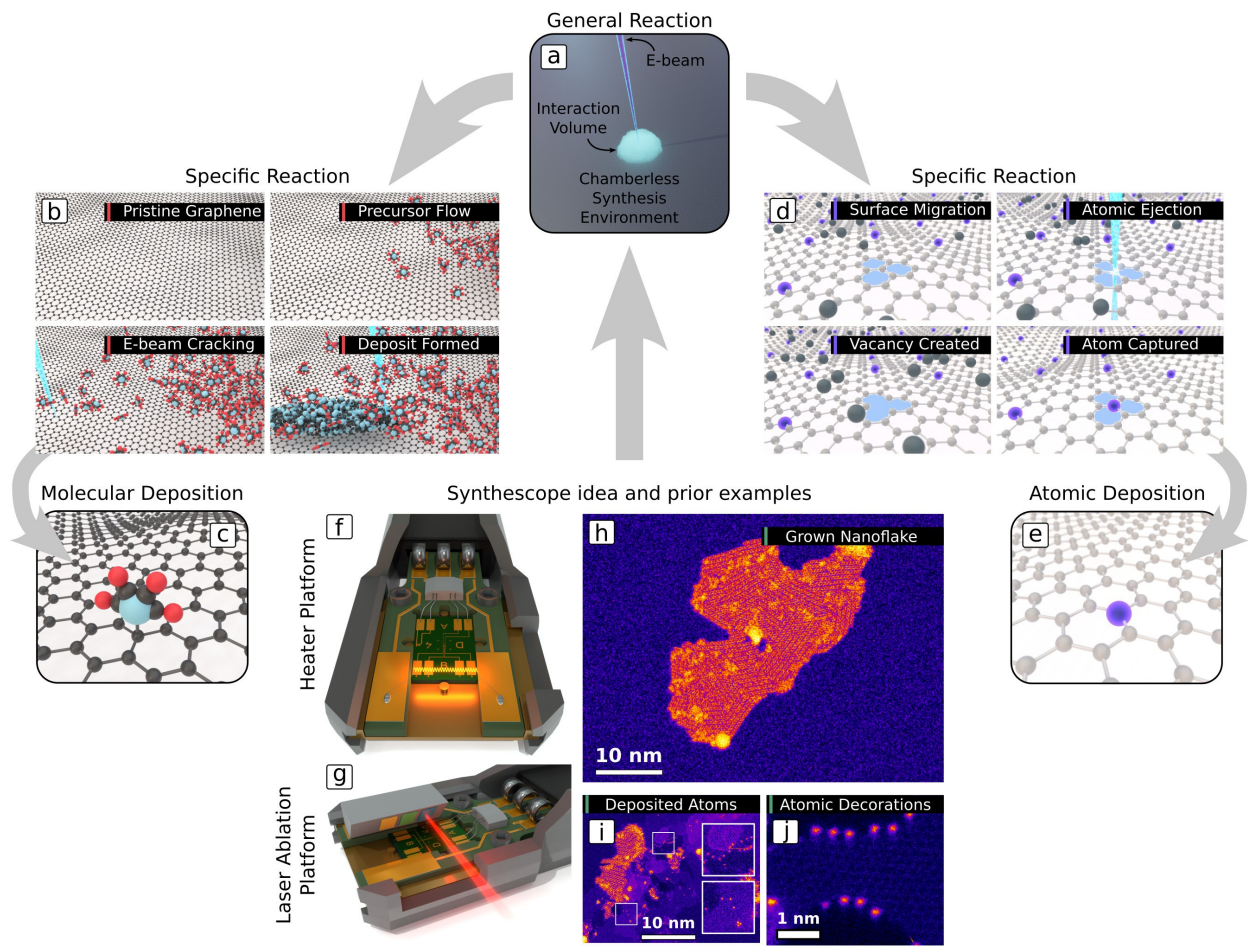

**Figure 1 Motivation for the synthescope idea.** (a) Depiction of a general, non-specific reaction driven by beam-sample interaction, conceived of as a chamberless synthesis environment (CSE). (b) Depiction of e-beam induced deposition as an example of a specific reaction—molecular dissociation (cracking) and bonding to the substrate. (c) Depiction of the deposition limit, a single molecule, or molecular fragment. (d) A different example reaction involving alteration of the substrate, instead of molecular dissociation, and subsequent capture of a diffusing adatom. (e) Depiction of the deposition limit, a single atom. (f) Depiction of a heater platform designed to deliver material to the sample in situ to support CSE reaction processes. (g) Alternate proposed method for material delivery via laser ablation. (h) Image of Sn nanoflake grown in situ on bilayer graphene using the heater platform. (i) Image of deposited single Sn atoms on bilayer graphene. (j) Image of evaporated Sn atoms decorating the edge of a graphene nanoribbon.

One of the envisioned embodiments of the synthescope idea is to use an *in situ* heater filament to deliver atomized material to the sample, Figure 1(f). In this conceptualization, the platform would maintain compatibility with existing electrical chips[18] to allow simultaneous operation of on chip heaters (e.g. commercially available chips or custom designed) and future electronic elements.

Experiments leveraging the use of such custom designed chips are critical for exploring electronic material properties and have underpinned discoveries of many kinds.[19–24] Likewise, incorporation of these types of custom-designed chips for operando use in the STEM has been successful as well.[10–12,25–27] A prototype implementation of our own design was presented elsewhere.[18] Using this platform[28] Sn nanoflakes were able to be grown on graphene, shown in Figure 1(h), and the deposition of single Sn atoms was recorded, shown in Figure 1(i). The Sn atoms were also introduced onto the edges of a milled graphene nanoribbon as shown in Figure 1(j). Subsequent tuning of the environmental parameters (i.e. substrate temperature) allowed for the direct writing of Sn atoms into a layer of graphene.[1]

While this approach shows promise, it also has drawbacks, one of which is the significant sample drift observed upon heating the filament. Thus, we also explore the second proposed system.

The second embodiment of the synthescope idea is to employ an *in situ* laser deposition system. The basic idea is to aim a laser at a target ablation block located close to and angled toward the sample. Atomized material, evaporated by laser irradiation, should impinge upon the sample and be observed during the deposition and growth process. While the abstract concept is simple, there are a number of considerations that must be attended to when implementing it on a real (and specific) microscope.

The first consideration is that, while the primary goal is an *in situ* laser deposition system, it would be remiss to ignore the other capabilities that a laser system could provide. For example *in situ* heating,[29–41] sample cleaning,[42,43] optical excitation, and ultrafast microscopy[44–49] experiments could be enabled through the use of the same (or similar) laser system if it is designed with this flexibility in mind.

In addition, when developing a solution to a problem, the solution is usually more valuable if it can be applied to other similar, or perhaps unforeseen, problems. Thus, a modular solution that can be generalized to other microscopes of the same type is sought, without requiring extensive or customized modification of the column or polepiece. Because we want to be able to image deposition and growth processes in real time with atomic resolution, the beam path cannot be blocked either before or after the sample while the laser system is in use.

Likewise, we envision incorporating deposition and growth processes with atomic fabrication workflows to enable integration with the custom designed experimental platforms, mentioned already, that will enable novel experimental capabilities based on the platform design. These platforms can be likened to the lab-on-a-chip approach of integrating measurement capabilities onto a chip but with the intention of scaling down beyond current lithographic patterning approaches to enable the exploration of spatially confined quantum structures, a quantum lab-on-a-chip (Q-LOC). To achieve this purpose, it is critical to facilitate, at least, electrical contact to the Q-LOC platform. This has already been demonstrated in several other publications leveraging existing equipment, but the non-trivial additional constraint this solution has introduced is that the sample cannot be tilted. Most laser systems that have been incorporated into STEMs require tilting the specimen toward the laser to provide a direct line of sight from the laser to the sample. Additionally, this strategy can be aided by a larger pole piece gap that allows the laser to be mounted above or below the sample by a significant margin; the larger the pole piece gap, the larger the angle of the laser system, and the smaller the tilt required to obtain a direct path. However, 1) larger pole piece gaps come at the cost of potentially reduced resolution, 2) ideally, we do not want to tilt the sample, and 3) we have already indicated that we would like to avoid extensive modification of the microscope hardware. Thus, we seek a solution that accommodates both zero tilt and the existing pole piece gap.

The desired solution should not interfere with the standard operation of the microscope when not in use. We do not seek to create a dedicated instrument that serves a singular purpose but to expand the capabilities of the microscope while retaining all other characterization and operating modes, thus preserving the existing flexibility. The sample vacuum chamber operates in the low $10^{-9}$ torr range. To retain this vacuum level, we should avoid using any O-ring seals. Likewise, keeping in mind future applications, we would also like to be able to easily interchange the probes to facilitate imaging, spectroscopy, or lasers designed for different purposes with minimal instrument downtime.

Here, we present the design and initial testing of an *in situ* STEM laser system (Vesta™ Optical Accessory, Waviks, Inc. Dallas, TX) that satisfies all of these requirements and is conceptually motivated by the use as a laser ablation/deposition system as depicted in Figure 1(g) but that can also be used for laser heating of the sample. Additionally, we discuss challenges and drawbacks of the design, particularly with regard to laser alignment and aiming, and our approach toward overcoming these challenges.

**Results and Discussion**

*System Design*

To satisfy the constraints mentioned in the foregoing discussion, we adopt the strategy presented in Figure 1. Briefly, a vacuum envelope, labeled as the Stationary Optics Tube (SOT), is mounted to an access port in horizontal alignment with the microscope (Nion UltraSTEM 200) stage. The laser is coupled to an optical fiber that feeds into the optical probe shaft with a short (typically 15-25 mm) focal length objective lens. The laser sources include an Oxxius 785 nm 350 mW single mode laser and a Waviks High Power Laser Controller with a 915 nm 25 W multimode laser. The optical probe shaft is mounted on a programmable nanomanipulator allowing control over the x, y, and z position. With this design, the majority of the optic elements and moving parts of the laser system remain at atmospheric pressure without requiring flexible vacuum seals. The optical probe shaft can be unmounted from the microscope and removed for bake-out and replaced without disturbing microscope vacuum. The only optic elements of the laser system within the vacuum chamber are the sapphire vacuum window and a mirror mounted to the sample holder. This mirror allows the laser to be reflected down onto the sample without requiring that the sample be tilted toward the laser to obtain a direct optical path. Different shapes or materials could be employed for the mirror, but here we will focus on a simple planar mirror.

The mirror is mounted to the metal base plate of a custom designed insert that interfaces with the existing electrical cartridge.[28] The lower panel in Figure 2 illustrates the laser reflecting from the mirror onto the sample, coincident with the electron beam. Importantly, this enables concurrent laser irradiation, atomically resolved imaging, and electrical characterization without tilting the sample.

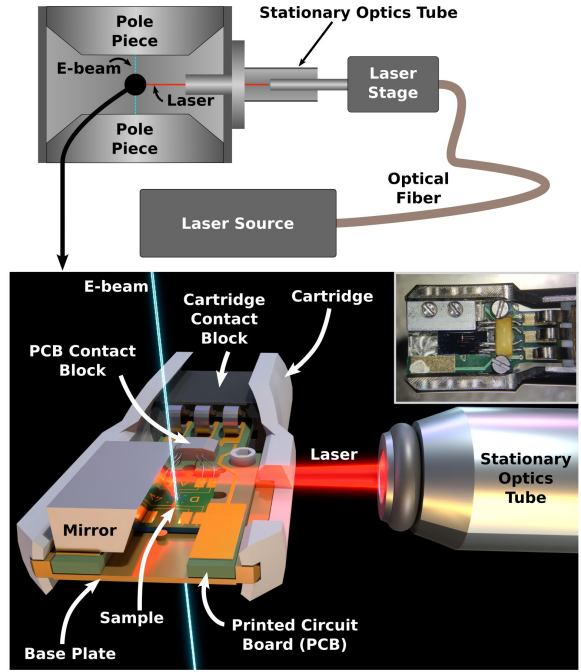

**Figure 2 Diagram of the side entry strategy.** The laser source is coupled to an optical fiber that terminates in an optical probe shaft mounted to a nanomanipulator. The optical probe shaft extends into the stationary optics tube (SOT) allowing close proximity to the sample stage while remaining at atmospheric pressure. Once the laser exits the optical probe shaft it is in free space, transmits into the vacuum via a sapphire window in the end of the SOT, and reflects off the mirror to finally arrive at the sample. A top-down photograph of the cartridge assembly is shown inset in the lower panel.

*Alignment*

With this design, the mirror is physically attached to the sample cartridge. Thus, moving the sample stage also moves the mirror, adding complexity to alignment of the laser and the e-beam. To simplify the discussion we will start by assuming that the microscope stage axes are well aligned to the laser axes (we will later relinquish this assumption). In this case, microscope stage movements can be compensated by laser movements as illustrated in Figure 3. Figure 3(a) shows a diagram of the geometry of the system and the main components; e-beam, laser, mirror, and sample. Figure 3(b) illustrates movement of the microscope stage along the x axis. The diagram is viewed along the same direction as shown in Figure 3(a) but the elements have been simplified to reduce visual clutter. Movement of the microscope stage in the -x direction (to the left on the page) is shown with light gray dashed-dotted sample and mirror elements.

Movement in the x direction (to the right on the page) is shown with light gray continuous sample and mirror elements. The original position is shown in black.

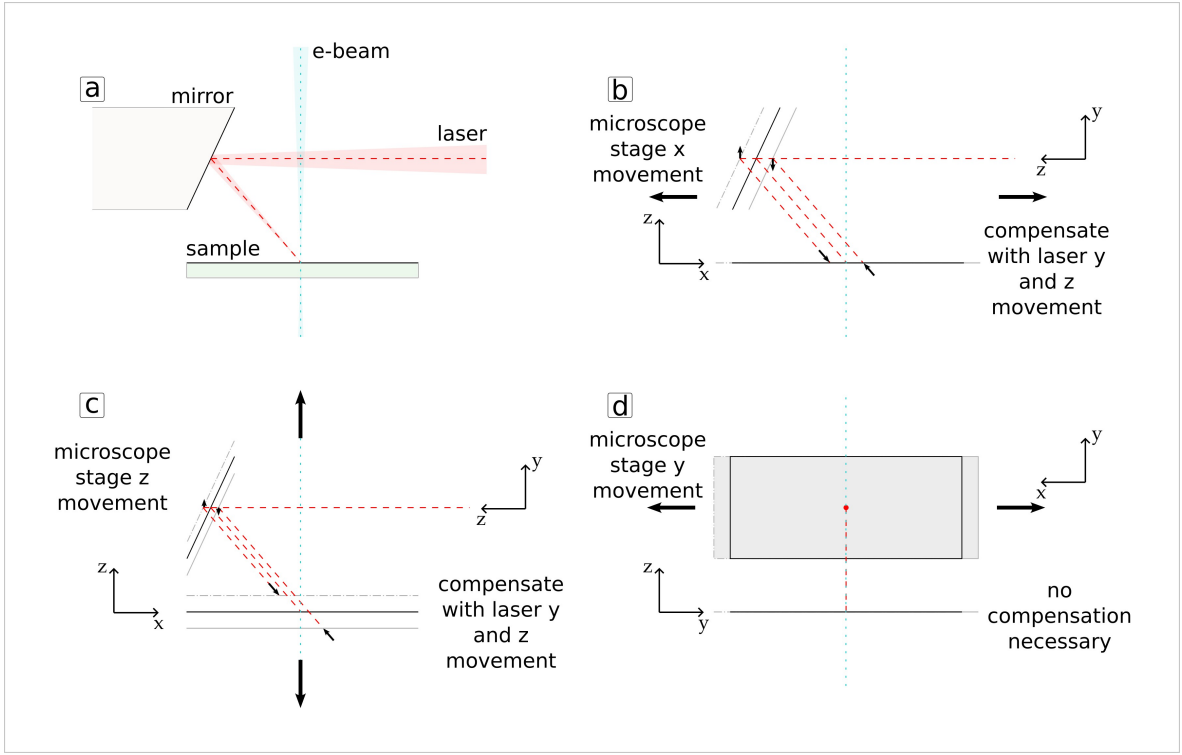

**Figure 3 Diagram of microscope stage movements and the effect on laser alignment.** (a) Diagram of the laser, e-beam, mirror, and sample. Subsequent diagrams are simplified to reduce visual complexity. (b) Diagram of microscope stage x movement, indicated by the large arrows. Small arrows indicate the adjustment needed in laser z and y directions to compensate for the stage movement. (c) Diagram of microscope stage z movement, large arrows. Small arrows indicate the adjustment needed in laser z and y to compensate. (d) Diagram of microscope stage y movement. No adjustment of the laser stage is necessary.

In this view, the position of the e-beam is effectively fixed, and the location is typically determined by the optical center of the objective lens. The alignment problem is then to bring the laser focal position to the same location.

The movement of the microscope stage elements causes a change in the laser optical path, illustrated by the different red dotted lines. This causes a misalignment between the laser position relative to the e-beam and focus relative to the sample surface. The small arrows indicate the compensating adjustment needed to bring the laser system back into alignment and focus. These adjustments involve both a laser stage z (focus) and stage y (vertical positioning) adjustment.

Similarly, in Figure 3(c) a diagram is shown of movement along the microscope stage z direction. Again, we find that laser stage compensation is needed for both the laser stage z and y to compensate for the microscope stage movement and bring the laser and e-beam back into alignment.

Figure 3(d) illustrates the microscope stage y movement. Here, the diagram depicts the elements rotated by ninety degrees (around the vertical axis) and looking along the laser path at the mirror. Because the mirror is translationally symmetric along this axis of movement, this operation has no effect on the alignment.

An additional challenge is that the laser spot cannot be seen using the e-beam unless the laser is physically changing the sample. Under the conditions here, the e-beam does not interact directly with the laser. A 'brute force' alignment option is clearly possible: the laser may be activated and the power output increased until it produces a sample alteration (assuming the laser is already focused on the sample) but, unless it is also already aligned with the e-beam, this sample change will not be detected. With a 'brute force' alignment strategy, the laser stage position must be moved (blindly) until the sample change is observed by the e-beam. Such a procedure can be performed methodically until the alignment coordinates are discovered (again assuming that the laser does not go out of focus) with the drawback of undesired sample alteration. We sought a more efficient alignment procedure, the development of which is described in what follows.

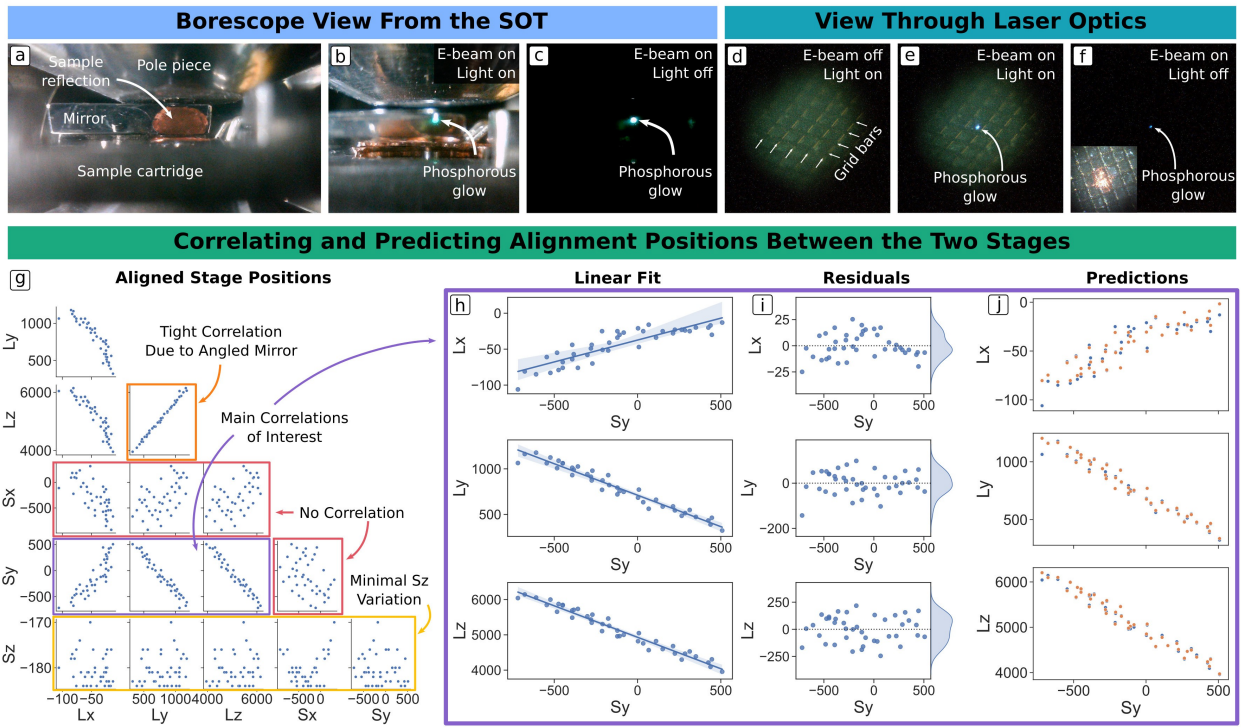

**Figure 4 Strategy for alignment.** (a) Borescope view from the stationary optics tube (SOT) looking into the pole piece chamber. The reflection of the TEM grid is visible in the mirror. (b) Borescope view of phosphor glow induced by the e-beam. (c) Borescope view of the phosphor glow with the pole piece light off. (d) Optical image of the sample looking through the laser optics. TEM grid bars are visible. (e) Phosphor glow produced by the e-beam. (f) Phosphor glow with the pole piece light off. Inset: laser spot on sample surface. (g) Pair plot of aligned positions. Sx, Sy, Sz represent microscope stage x, y, z and Lx, Ly, Lz, represent laser stage x, y, z, positions. All values are in microns. (h) Linear fit to the Sy correlations. The shaded region represents a 100% confidence interval. (i) Residuals of the linear fit with a kernel density estimate plot on the side. (j) Original alignment positions shown in blue and predicted positions shown in orange.

To improve this situation, we leverage the existence of the optics that are already used to guide and focus the laser onto the sample. The same optics can be used to collect an image of the sample as viewed by the laser. The Vesta™ optical probe includes high magnification imaging of the sample for focusing the laser spot. A semi-reflective beam splitter allows the laser spot to be imaged on the sample surface. With an available optical image the determination of laser focus becomes straight-forward. However, the position of the laser on the sample is still undetermined. The ideal position of the laser on the sample is dictated by the position of the e-beam, so we need a way to see the e-beam position using the laser optics. To accomplish this task, we sprinkled phosphor powder across the sample (SPI-Chem™ TEM Screen Recoating Phosphor Powder or alternatively

SPI-Chem™ P-47 Scintillator Powder for Recoating Scintillators). When the e-beam is positioned on a phosphor particle it glows brightly and can be seen in the laser camera.

Figure 4(a)-(f) shows a summary of how alignment positions can be found using phosphor particles and the laser optics. Figure 4(a) shows a borescope view into the pole piece chamber, looking through the window in the end of the SOT, with various features labeled. The borescope view is not using the laser optics, but a separate set of optics (requiring removal of the optical probe shaft), giving a wide, low-magnification field of view. Based on this work, a revised design of the optical probe shaft has been undertaken to integrate a second set of optics that allows simultaneous acquisition of a wide angle borescope view and the view from the laser optics. This allows easier coarse and fine positioning of the laser. This capability has been successfully deployed on another laser system but has not yet been integrated with the system described here.

As can be seen in Figure 4(a) and (b), the sample is not directly visible, as viewed by the laser, because the view port is approximately in the same plane as the sample. Instead the sample reflection is visible in the mirror fixed to the far side of the sample holder.

With the e-beam positioned on a phosphor particle, the glow can be observed in the mirror, shown in Figure 4(b). To confirm that we were not observing a spurious reflection from another light source, the pole piece light was turned off, shown in Figure 4(c).

Figure 4(d) shows a view of the sample using the laser optics. Because the laser is focused, we observe a magnified optical image of the sample. The grid bars of the sample are indicated by arrows. When the e-beam is positioned on a phosphor particle, shown in Figure 4(e), we can observe the emission. Likewise, activating the laser produces a visible spot on the sample where the laser is focused, shown in the inset in Figure 4(f). With this information we can easily log the position of the laser as seen by the camera, and move the laser stage so that the phosphor particle appears at the same position. The advantage here is that the laser need only be turned on once and the position logged. The rest of the alignment does not require the laser to be activated, thus preserving the sample. Additionally, because we are observing reflected light, we can use a low laser power that does not alter the sample. Figure 4(f) shows an image of the glow of a much smaller phosphor particle that can easily be seen with the pole piece light turned off.

Even with this much improved alignment strategy, large microscope stage movements can result in a shift that is greater than the field of view of the laser optics or a significant focus change. Ideally, we would be able to move the microscope stage to an arbitrary position and, based on the microscope stage values, predict the new laser stage position that would bring it into alignment. If this can be done, it would provide the first step toward automating the movement of the laser stage so that it perpetually maintains alignment.

To accomplish this task, we assume there exists a linear relationship between the two stages such that each microscope stage (x, y, z) position uniquely determines the required laser position. Let $S_x$, $S_y$, and $S_z$ represent the microscope stage x, y, and z positions, respectively. Let $L_x$, $L_y$, and $L_z$ represent the laser stage x, y, and z positions, respectively (x, y, and z for the laser being in different directions from the x, y, and z for the microscope stage). Assuming a linear relationship we may express $L_x$ as proportional to some combination of the microscope stage values:

$$L_x = T_{xx}S_x + T_{xy}S_y + T_{xz}S_z + C_x$$

The values $T_{nm}$ are simply constants of proportionality that we will need to find experimentally. Likewise, $C_x$ is a constant offset that must also be determined experimentally. In the same manner, we can express $L_y$ and $L_z$ as functions of the microscope stage parameters with their own constants of proportionality:

$$L_y = T_{yx}S_x + T_{yy}S_y + T_{yz}S_z + C_y$$

$$L_z = T_{zx}S_x + T_{zy}S_y + T_{zz}S_z + C_z$$

Recasting this system of equations into matrix notation we may write:

$$L = \boldsymbol{T}S + C$$

Here, $L$ is our laser stage vector composed of $L_x$, $L_y$, and $L_z$. $S$ is our microscope stage vector composed of $S_x$, $S_y$, and $S_z$. $\boldsymbol{T}$ is a 3x3 transformation matrix, composed of all the $T_{nm}$ values. We can then fit an equation of this form to experimentally acquired alignment data and obtain a least squares estimate of the parameters. The code used to perform this calculation, as well as to reproduce Figure 4(g)-(j) can be found at the GitHub repository https://github.com/ondrejdyck/Side-Entry-Laser-Alignment.

Alignment positions were found across the sample and are presented in the pair plot shown in Figure 4(g), which shows a scatter plot of the variation of each coordinate with each other coordinate. A few features are worth pointing out: (1) We observe that $L_y$ is tightly correlated with $L_z$. This is because $L_y$ changes the vertical position of the laser on the mirror which sits at an angle and changes the distance to the sample, requiring an adjustment in $L_z$. The microscope stage positions $S_x$ and $S_y$ are, by definition, uncorrelated. Nevertheless, the non-random manual sampling of positions across the sample could produce an artificial correlation. However, because we know these are uncorrelated we can use this pattern of data points as a representation of what uncorrelated data will look like. Based on this insight we can see that $S_x$ is not correlated with $L_y$ and $L_z$ and perhaps only slightly correlated with $L_x$. Since all the data shown here was acquired with the sample in focus, with respect to the electron beam, fluctuations in $S_z$ represent only small changes in sample height, spanning about 15 μm. As we will show, this is beyond the precision of alignment and the scatter here is simply noise.

The parameters we are most interested in are the correlations between $S_y$ and the laser stage positions. As the microscope stage moves toward or away from the laser stage we have substantial adjustments that need to be made to maintain the alignment and therefore a substantial correlation exists between these parameters. Before finding our transformation matrix we can use these correlations to get a sense of the alignment uncertainty in the $L_x$, $L_y$, and $L_z$ parameters. First, we perform a linear fit to the data, shown in Figure 4(h). Then, we calculate the residuals, shown in Figure 4(i). The plot of the residuals gives us a measure of the repeatability of the alignment process. Along the right side of the scatter plots we show a kernel density estimate to give a sense of the distribution. We find that the precision of $L_x$ is +-20 μm, that of $L_y$ is +-100 μm, and that of $L_z$ is +-200 μm. Very likely, the precision of $L_x$ is an underestimate due to the fact that the x position of the e-beam does not change and so $L_x$ is simply sampling the same location repeatedly. In any case, we see that the 15 μm variation in sample height is negligible compared to the uncertainty in the other parameters.

By fitting an equation of the form described above we can generate a least squares estimate of the linear transformation required to convert microscope stage values to laser stage values. Since we already know what the recorded alignments are we can check how well this strategy works. The original alignment values and the predicted alignment values are compared in Figure 4(j). The blue datapoints represent the measured alignment values and the orange datapoints represent the predicted alignment values. In a perfect world, these would coincide at every location. The result is encouraging because we see that the deviation in predictions is no greater than the original spread in precision.

With this strategy we were able to navigate to arbitrary microscope stage positions and compensate with predicted laser stage adjustments such that the glowing phosphor particle could be seen through the laser optics after every movement. Many of the alignment data points shown in Figure 4 were first roughly aligned using laser stage values predicted from microscope stage positions. Fine adjustments were then handled manually by observing the position of the phosphor particle using the laser optical camera. Further optimization would likely require a higher precision nanomanipulator. Despite the added alignment complexity of introducing an optical element that moves with the sample (the mirror), the alignment process can be significantly simplified by leveraging various optical views of the sample and phosphorous particles that reveal the e-beam location and the position of the laser without the laser even being activated.

*Examples of Use*

So far, we have focused primarily on the motivations behind the side entry laser system, the physical constraints of such a system, our implementation and emergent challenges and solutions with regard to alignment. Having dealt with these topics at some length, it would be remiss not to present some examples of the laser system in use. Here, we show the laser system being used to clean a graphene sample across a wide area, Figure 5(a), as well as to induce a phase change in a $PdSe_2$ flake, Figure 5(b). We do not intend to delve into great detail regarding the physics of these processes. The main purpose is to illustrate that the laser system works as intended, as far as delivering heat to a sample is concerned. Sample preparation is detailed in the supplemental materials.

For both examples the microscope was operated at 80 kV accelerating voltage. In the graphene example the laser was activated at 100% power continuous duty cycle for less than a second (manually turned on and then off). We observed a pronounced cleaning effect that has been reported in the prior literature.[42] Subsequent images of the surrounding area were acquired at lower magnification moving away from the central region extending to about 20 μm away. High resolution images of the graphene (the image shown appears to be a triple layer region) could be acquired at room temperature without observing hydrocarbon deposition. This should be contrasted with results employing the use of a heating chip, where visible contamination is removed but hydrocarbon deposition is still present.[50] In that work, high temperatures were maintained and hydrocarbon deposition was controlled/mitigated by depositing barriers that blocked further ingress of contaminants. Here, and in the prior literature exploring laser cleaning,[42] it appears that the cleaning effect is superior to that of a heating chip. Likely this is due to a much hotter temperature being achieved than is available with a heating chip.

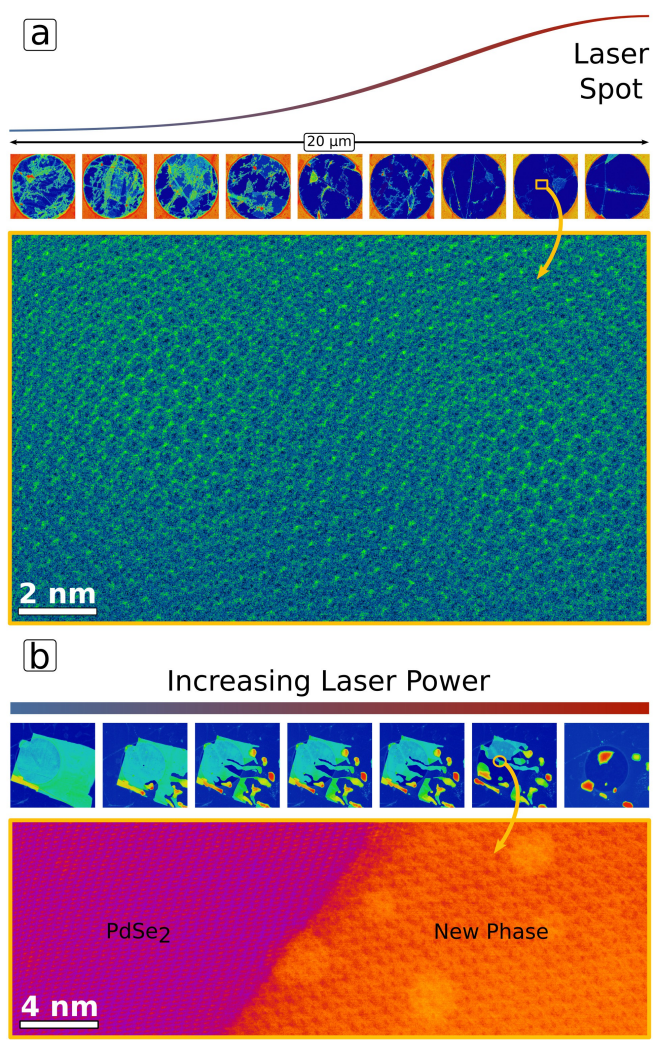

**Figure 5 Examples of laser heating.** (a) Laser cleaning of graphene using 100% power. (b) Phase change driven by laser heating in a PdSe$_2$ flake. Incremental increases in laser power were used.

In the second example, shown in Figure 5(b), we show a PdSe$_2$ flake that has received a stepwise increase in laser power. Again, we operated the laser manually with a continuous power output, simply turning it on and then off. As the laser power was increased we were able to drive a phase change in the PdSe$_2$ flake, the details of which are still under investigation. Each image was acquired with the laser off and the sample at room temperature. In this way, we can drive a phase change gradually, freezing the change in place for further examination.

*Laser Evaporation*

As we have already mentioned in the introduction, the overarching goal of developing this laser system is to enable the supply of foreign material to the sample while concurrently maintaining the rest of the microscope capabilities. This strategy, it is conjectured,[17] will enable controlled delivery of atomized material to the sample, allow the atomically-resolved observation of synthesis processes, and possibly their control, if environmental parameters can be adjusted to produce reactions within the beam-sample interaction volume—the CSE. So far, we have detailed the challenges encountered when attempting to align the laser to the e-beam location, a difficulty which is compounded by the movement of the mirror with the microscope stage. We have showed how this difficulty may be addressed and provided examples of the laser in use (aligned with the e-beam). However, this use of the laser was not the primary motivation for the described laser system.

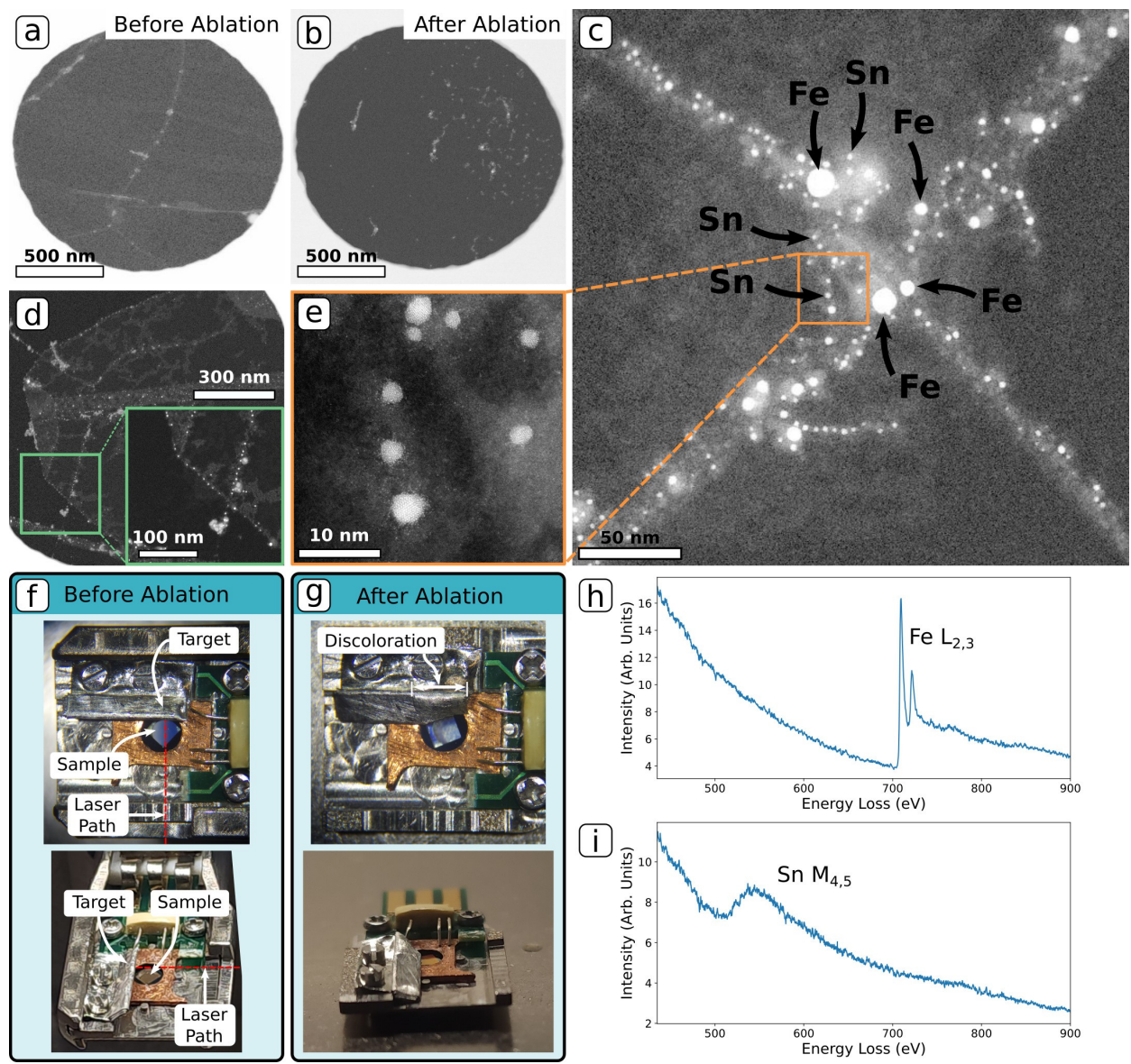

**Figure 6 Summary of laser-induced deposition.** (a) Representative HAADF image of suspended graphene prior to laser ablation attempts. (b) Representative HAADF image of suspended graphene after laser ablation attempts. (c) Magnified view of surface nanoparticles with examples of several Sn and Fe particles pointed out. (d) Image of a graphene step edge after laser ablation. Sn nanoparticles were observed at the step edge. Inset shows a magnified view. (e) Magnified view of the boxed area in (c). (f) Optical images of the custom insert that facilitates the attachment of both the laser mirror and laser target. In the configuration pictures, the mirror has been removed and replaced with a Sn foil target. (g) Optical images of the custom insert after the laser ablation attempts. The Sn foil target is visibly discolored in one region and has drooped down onto the sample. (h) Representative Fe EELS signal acquired on one of the particles marked "Fe" in (c). (i) Representative Sn EELS signal acquired on one of the particles marked "Sn" in (c).

To test the laser system for its intended purpose, the mirror was replaced with a 3D printed backer block with approximately the same shape as the mirror. Al foil was wrapped around the backer block and laser evaporation of the foil attempted. The initially installed (Oxxius 785 nm 350 mW) laser did not have sufficient power to evaporate the foil, likely due to its thickness. To address this problem, a more powerful laser (Waviks High Power Laser Controller with a 915 nm 25 W multimode laser) was installed. Notably, because of the flexible design of the laser system, this process did not involve breaking vacuum. Moreover, the optical probe shaft is secured to the nanomanipulator using just four bolts, making removal and replacement simple.

The high power laser was able to evaporate Sn foil (Xiamen Futiantian Technology Co., Ltd, >99.99% purity, 0.05 mm thickness) wrapped around the backer and subsequently evaporate the backer material itself, see supplemental material. This demonstrates *in situ* evaporation of the Sn foil, however, for our intended application we are interested in deposition onto the sample, not simply evaporation.

The experimental set-up was revised by removing the backer block and creating a target by folding the Sn foil as shown in Figure 6(f). With this design, laser evaporation of the Sn target and deposition of Sn nanoparticles onto the sample was achieved.

A graphene specimen, supported on a perforated silicon nitride membrane (Silson Ltd.), was used as the deposition substrate. The suspended graphene areas, shown in Figure 6(a), provide a material with a low high angle annular dark field (HAADF) signal that is helpful for allowing the detection of heavier atoms and nanoparticles that may be evaporated onto the surface, which will appear much brighter.[51] As is typical,[42,52–56] the graphene membrane harbors residual surface contaminants and, in particular, Fe nanoparticles that are likely residue from the graphene transfer process.

During attempts at evaporation, many laser positions and parameters were explored which will not be discussed here because we were unable to achieve deposition at any location that was simultaneously being imaged. Nevertheless, after these evaporation attempts, we were able to locate some areas of the sample that contained a conspicuously larger number of nanoparticles adhered to the surface, one of which is shown in Figure 6(b). A magnified view of one of these locations is shown in Figure 6(c) and there appears to be a bimodal distribution of nanparticle sizes; some distinctly larger and some distinctly smaller. Using electron energy loss spectroscopy (EELS) the characteristic fingerprint of Fe ($L_{2,3}$ EELS edge) and Sn ($M_{4,5}$ EELS edge) could be clearly identified for the larger and smaller particles respectively. Representative EELS edges are shown in Figure 6(h) and (i).

A further magnified view of some of the smaller nanoparticles is shown in Figure 6(e). A quantitative analysis of nanoparticle size was not undertaken but they appear to be roughly in the range of 2-5 nm in diameter with evidence of single separated atoms in the surrounding material (EELS was not attempted on these isolated atoms so it is premature to conclude with certainty that they are Sn, but this seems likely given the evidence shown).

Another area containing a portion of bilayer graphene was also examined, shown in Figure 6(d). Here, a preponderance of Sn nanoparticles were observed decorating the step edges suggesting that surface migration is simultaneously taking place. A magnified view of some of these particles is shown in the inset.

After removal of the cartridge from the microscope, additional images of the sample platform with the laser target were obtained, shown in Figure 6(g). The Sn target has a clearly discolored region and the target has collapsed over the sample, suggesting that it was close to melting during laser irradiation.

Further optimization of the target and evaporation parameters are clearly warranted, to enable highly controlled deposition in precisely the desired locations. Nevertheless, the aim of achieving *in situ* laser-induced deposition onto a sample has been accomplished, which is the initial significant step toward these later refinements.

*Discussion*

*Alternative Laser Positions*

A thorough review of alternate strategies for laser illumination in a TEM has recently been given by A. M. Zak.[57]

In this work, we have chosen to mount the laser system so that it occupies an existing access port, positioned at ninety degrees relative to the sample. This laser position, coupled with a non-tilting sample holder has led to the requirement for the mirror mounted to the sample stage, which complicates the alignment process. In particular, microscope stage movement can disturb the laser-e-beam alignment.

An alternative laser position (not included in the discussion of A. M. Zak) would be to mount the laser so that is shines along the e-beam optical pathway propagating parallel to the electrons. In principle, one could insert a laser where the e-beam makes a sharp change of direction, such as in a spectrometer or monochromator, and mount the laser there. Once the alignment is obtained, movement of the microscope stage should not affect the laser-e-beam alignment.

This design has some obvious drawbacks, however. (1) This would require significant modification of the microscope hardware. (2) It could require a long laser focal length. (3) The laser would not be able to be used for ablation of an adjacent target. (4) It is possible the laser in could propagate to other elements of the column.

*A Tilting Holder*

The experiments shown here have employed the use of a fixed sample holder, without tilt capabilities. In exploring tilting options, we have determined that our existing double-tilt holders, with maximum tilt angles of about 20 degrees, do not allow a direct line-of-sight pathway to the sample from the laser. Thus, if tilting capabilities are to be developed, they must involve other modifications or also employ a mirror.

One of the challenges with a tilting holder that also harbors a mirror is that additional laser compensating movements will be necessary to maintain alignment during tilting. As the sample is tilted (unless one happens to be looking at the eucentric point) there will be a change in height, which could be compensated by a vertical stage movement to focus the e-beam. This stage movement will change the mirror position and require a laser adjustment. The predictive equations developed above would need to be modified to account for these additional degrees of freedom.

These challenges do not seem insurmountable and we hope to explore this option in future work.

*Stability*

Questions of stability naturally arise when adding complex components to a sensitive instrument. To help reduce instabilities, the microscope housing (designed to reduce ambient mechanical vibrations) had to be modified/expanded so that the laser stage fits within this housing. During operation of the laser, after alignment, we have not observed drift in the alignment due to laser activation or sample heating. It is likely that the laser spot size, which is on the order of ten microns, is simply much too large to drift significantly out of alignment when the field of view of the electron beam is typically less than two microns.

Nevertheless, there is a change in height that is significant, with respect to e-beam focus, when the sample is heated. This height change is attributed to thermal expansion of the support membrane. To compensate for this without moving the stage (which moves the mirror) the 'fine focus' can be used, which alters the microscope accelerating voltage to achieve a different focal plane without changing the objective lens excitation.

During the ablation experiments, we also observed substantial movement of the sample, which suggests significant transfer of the laser power onto the sample. We are exploring further optimization of the target geometry and surface quality to reduce this effect.

*Integration with other In Situ Techniques*

The design presented here has been built around maintaining compatibility with the electrical contacting cartridges that allow, for example, biasing and transport measurements to be performed *in situ*. All the experiments shown used this cartridge design and we, therefore, now have the capability to also integrate electrical experiments with this system. We expect that future design of on-chip measurement capabilities that can be electrically controlled (i.e. the Q-LOC platforms, mentioned previously) will enhance *in situ* capabilities.

The question arises, what other *in situ* capabilities might be integrated in the future with this system. We might ask whether we could integrate the laser capabilities with, say, liquid/gas-flow. With this specific instrument it is not possible. The stage is designed for deposited cartridges rather than side entry, making the supply of liquid/gas problematic. Nevertheless, a similar laser system could be installed on a side-entry-style microscope to facilitate these kinds of experiments. While this is true, we expect that unwelcomed surprises would be in store for anyone who unwittingly evaporates the thin membranes that contain the liquid/gas or who inadvertently converts their liquid flow to a gas flow.

Moreover, the liquid/gas cell strategy supplies macroscale quantities of materials. For the atomic-scale fabrication processes we have envisioned, we would like the ability to supply very minute amounts of materials, down to single atoms. It is for this purpose that have introduced the laser ablation idea. Once this system has been optimized, we believe the user will be able to rapidly switch between the deposition of multiple different source materials from multiple different targets. Combined with e-beam modifications such as sculpting[17], writing[1], patterning[7], and seeded growth[28], we expect a vast landscape of opportunities to explore.

**Conclusion**

In this work we have argued that the trend in atomic manipulation using electron beams is a step towards incorporating synthesis processes into the electron microscope. We suggest that these processes will be a significant driving force for enhanced fabrication capabilities in the future. The challenge will be to provide integrated solutions capable of delivering materials for atom-by-atom synthesis and capable of providing independent controls of the local environment.

Here, we have implemented a design where we use a mirror that reflects the laser onto the sample. This system is relatively inexpensive and can be incorporated into an existing STEM with minimal modifications. We have presented an examination of the alignment challenges this system presents as well as strategies to overcome these challenges. We showed simple, relevant examples of the laser being used to heat 2D materials: (1) cleaning a bilayer graphene sample and (2) inducing a phase transformation in $PdSe_2$. The mirror can easily be replaced with an ablation target to convert the system from a laser irradiation system to a laser ablation system. For the ablation process to be successful on bulk materials—as opposed to the electron transparent samples—a higher power laser is also required. A Sn foil target was used to evaporate Sn nanoparticles onto a graphene sample as a proof of principle.

The results presented describe the development of an *in situ* laser system that can be used for atomic scale synthesis and processing. We hope that these results will help to guide the field in productive new directions, inspiring improvements, revisions, and new approaches toward enhanced capabilities.

**Acknowledgments**

Electron microscopy and phase transformation was supported by the U.S. Department of Energy, Office of Science, Basic Energy Sciences, Materials Sciences and Engineering Division (O.D., K.X., A.R.L., S.J.) and was performed at the Oak Ridge National Laboratory's Center for Nanophase Materials Sciences (CNMS), a U.S. Department of Energy, Office of Science User Facility. The Vesta™ side entry laser accessory described here was developed and produced by Waviks, Inc. (Dallas, TX). Ben Wolf and Thomas Moore have equity interests in Waviks, Inc.

**Supplemental Information**

*Fabrication of graphene-PdSe$_2$-graphene samples on TEM grids*

First the few layer PdSe$_2$ flakes were exfoliated on SiO$_2$/Si substrate using the gold-assisted method[1] from a self-flux grown bulk crystal. Then the PdSe$_2$ flakes were picked up using poly propylene carbonate (PPC) film and transferred to the CVD-grown graphene on copper. Afterward, a separate graphene/Cu was spin-coated with poly propylene carbonate (PPC) at 3000 rpm, 50 seconds, etched with FeCl$_3$, and rinsed in DI water a minimum of three times. The PPC-graphene film was placed on the dry transferred PdSe$_2$/graphene/Cu and gently heated first at 60 °C for 10 minutes and then at 90 °C for 15 minutes to enhance adhesion between the PPC-graphene and PdSe$_2$-graphene-Cu. The PPC-graphene-PdSe$_2$-graphene-Cu was placed gently in FeCl$_3$ solution to etch away the bottom copper, leaving PPC-graphene-PdSe$_2$-graphene. Finally, the PPC-graphene-PdSe$_2$-graphene was transferred on the TEM grid and annealed at 100 °C for 5 minutes to ensure good adhesion. The top PPC was removed by dissolving with chloroform for 2 hours. Figure S1 shows scanning electron microscope images of the graphene sandwiched PdSe$_2$ flakes.

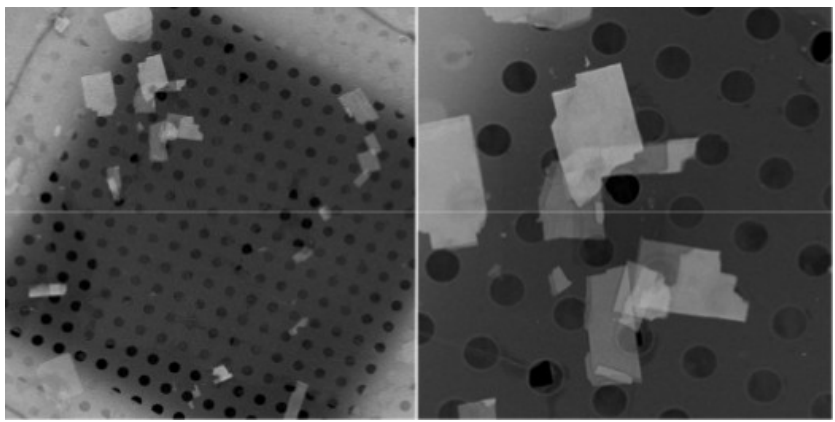

**Figure S1** SEM image of the graphene sandwiched PdSe2 flakes

*Initial Laser Deposition Attempt*

In an initial attempt at laser-induced evaporation and deposition, a single layer of Sn foil wrapped around a 3D printed backer block was used, as illustrated in Figure S2(a). During evaporation attempts the Sn foil was able to be melted/ablated as shown in Figure S2(b). Folding and drooping of the Sn foil indicates the extent of the heated region. Figure S2(c) shows the piece of Sn foil unwrapped from the backer block. Figure S2(d) shows the holes in the backer block produced through laser irradiation. While we can conclude that some material was evaporated, we did not observe deposition onto the sample.

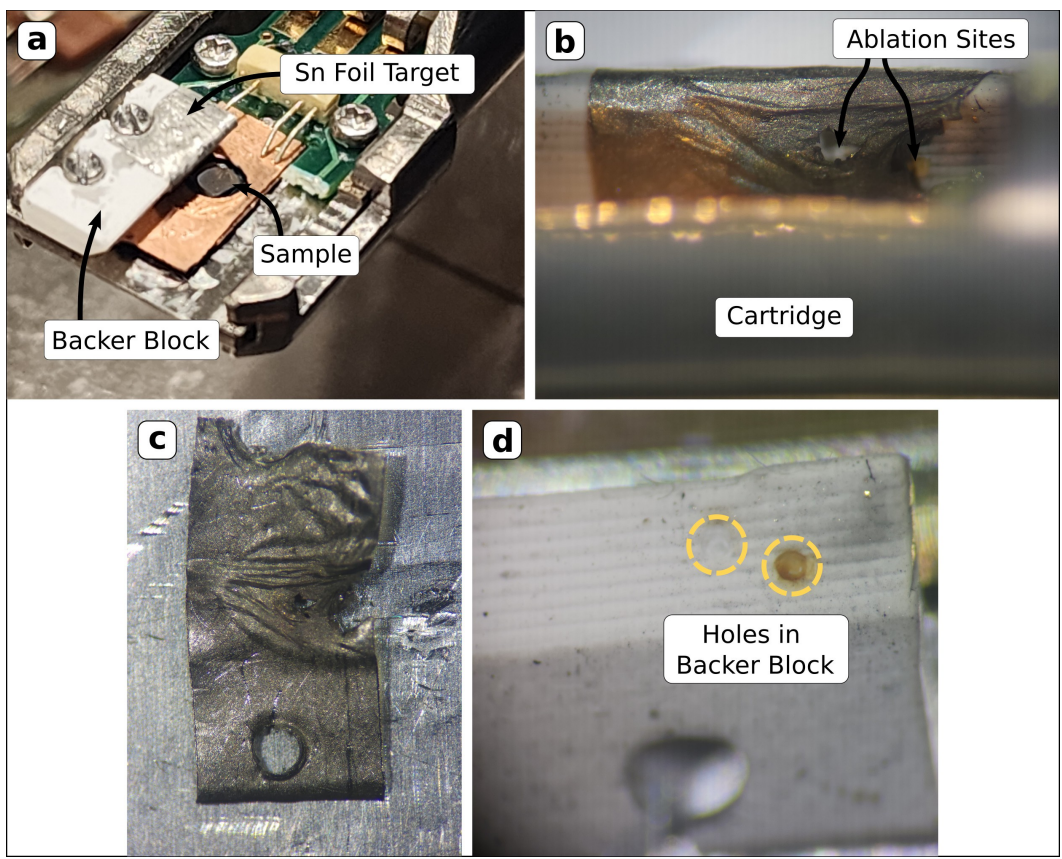

**Figure S2 Summary of initial deposition attempt.** (a) Initial configuration prior to laser irradiation. (b) Optical image of the face of the Sn foil after laser irradiation. (c) Sn foil unwrapped from backer block. (d) Backer block material has been ablated.

Geohegan, D. B.; Xiao, K. PdSe2: Pentagonal Two-Dimensional Layers with High Air Stability for Electronics. *J. Am. Chem. Soc.* **2017**, *139* (40), 14090–14097. https://doi.org/10.1021/jacs.7b04865.